\documentclass[journal]{IEEEtran}

\usepackage{amsmath,amssymb,amsfonts}
\usepackage{soul}
\usepackage{pifont}
\newcommand{\xmark}{\ding{55}}%
\usepackage{enumerate}
\usepackage{colortbl}
\usepackage{algorithm} 
\usepackage{algpseudocode}
\usepackage{graphicx}
\usepackage{textcomp}
\usepackage{tikz,caption}
\usepackage{pgfplots}
\pgfplotsset{compat=1.17}
\usepackage{xcolor}
\usepackage[inline]{enumitem}
\usepackage{float} 
\usepackage{cite}

\newlength{\TenSpaces}
\ifCLASSINFOpdf

\else

\fi

\begin{document}

\title{\huge \textbf{HOACS}: Homomorphic Obfuscation Assisted Concealing of Secrets to Thwart Trojan Attacks in COTS Processor}

\author{\IEEEauthorblockN{Tanvir Hossain, Matthew Showers, Mahmudul Hasan, and Tamzidul~Hoque
}
\thanks{\IEEEauthorrefmark{1} Tanvir Hossain Matthew Showers, Mahmudul Hasan, and Tamzidul~Hoque are with 
the Department of Electrical Engineering and Computer Science, The University of Kansas, KS 66045, USA (e-mail: tanvir@ku.edu, mds2016@ku.edu, m.hasan@ku.edu, hoque@ku.edu). }}

\markboth{Journal 2024}%
{Hossain \MakeLowercase{\textit{et al.}}: \textit{HOACS}: Homomorphic Obfuscation Assisted Concealing of Secrets to Thwart Trojan Attacks in COTS Processor}

\maketitle

\begin{abstract}

Commercial-Off-the-Shelf (COTS) components are often preferred over custom Integrated Circuits (ICs) to achieve reduced system development time and cost, easy adoption of new technologies, and replaceability. Unfortunately, the integration of COTS components introduces serious security concerns. None of the entities in the COTS IC supply chain are trusted from a consumer's perspective, leading to a ``zero trust” threat model. Any of these entities could introduce hidden malicious circuits or hardware Trojans within the component, allowing an attacker in the field to extract secret information (e.g., cryptographic keys) or cause a functional failure. Existing solutions to counter hardware Trojans are inapplicable in such a zero-trust scenario as they assume either the design house or the foundry to be trusted and consider the design to be available for either analysis or modification. In this work, we have proposed a software-oriented countermeasure to ensure the confidentiality of secret assets against hardware Trojans that can be seamlessly integrated in existing COTS microprocessors. The proposed solution does not require any supply chain entity to be trusted and does not require analysis or modification of the IC design. To protect secret assets in an untrusted microprocessor, the proposed method leverages the concept of residue number coding (RNC) to transform the software functions operating on the asset to be fully homomorphic. We have implemented the proposed solution to protect the secret key within the Advanced Encryption Standard (AES)  program and presented a detailed security analysis. We also have developed a plugin for the LLVM compiler toolchain that automatically integrates the solution in AES. Finally, we compare the execution time overhead of the operations in the RNC-based technique with comparable homomorphic solutions and demonstrate significant improvement.   
\end{abstract}


\IEEEpeerreviewmaketitle

\section{Introduction}
To meet the increasing performance and unique functionalities, systems developers opt for custom-designed hardware, such as Application Specific Integrated Circuits (ASICs). Design and fabrication of such components require significant financial investment and greater time-to-market. Commercial-off-the-shelf (COTS) components, on the other hand, are easier to adopt as they do not require any involvement in the design and fabrication process of the Integrated Circuits (ICs). Hence, COTS components offer lower system development costs, shorter time-to-market, and easier integration of emerging IC technologies~\cite{cots3}.  Currently, COTS components are commonly used in diverse applications, including military, communication, automotive, avionics, and commercial electronic systems. According to data published by the Department of Defense (DoD) in 2014, 72\% of their ICs were COTS \cite{shanahan2014department}. The substantial presence of COTS components is also common for electronic systems developed by the Australian Military \cite{cots1}.

\begin{figure*}[t!]
    \centering
    \includegraphics[width=\linewidth]{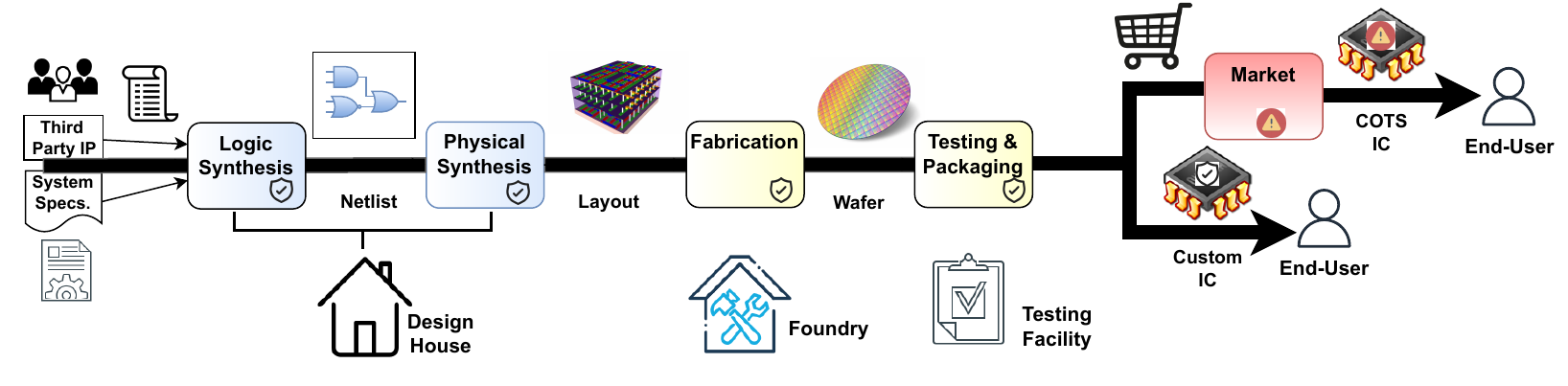}
        \caption[Custom IC Design Process]{The common stages of designing a custom IC, 
        compared to the process involved in acquiring COTS IC.
        For a consumer of the COTS IC, none of supply chain stages are accessible and all involved vendors are assume to be untrusted.
    }
    \label{fig:ch1_customIC}
    \vspace{-0.4cm}
\end{figure*}

While utilizing COTS parts can provide many benefits, it also introduces serious security concerns. Due to the complete reliance on third parties for design and fabrication, the threat of malicious modification or hardware Trojan insertion is far greater compared to in-house custom ASIC \cite{ht_bhunia}. Hardware Trojans are hidden malicious alterations to the IC design that could threaten the system's confidentiality and integrity. The threat model of COTS hardware can be considered as ``Zero Trust''; where none of the supply chain entities, including any design, fabrication, and test vendors can be trusted.  On the other hand, as shown in Fig. \ref{fig:ch1_customIC}, for custom ICs with in-house design, the design stage could be trusted and the subsequent supply chain stages remain untrusted. For instance, while developing a mission-critical electronic system a military entity could purchase all components  off-the-shelf or could design the components by themselves (i.e., custom IC). In the latter case, at least the design of the IC is trusted. If the military entity has access to a trusted foundry, the fabrication process could be trusted as well. Therefore, choosing COTS component to design the system could be an easier option to adopt but involves increased hardware Trojan threat as none of the supply chain entities can be trusted. The recent enactment of the CHIP and Science Act 2022~\cite{whitehouseFACTSHEET} from the US government aims to strengthen domestic electronics manufacturing capabilities. Despite the advantages of in-house manufacturing, there remains the potential for adversaries within the facilities themselves. Moreover, from a global perspective, COTS chips are extensively used, and authorities worldwide must address the security implications of such components.    

Existing research on hardware Trojans primarily focuses on addressing Trojan insertion in two major supply-chain entities: untrusted foundry and untrusted hardware intellectual property (IP) vendors. 
Techniques to detect Trojans inserted in an Integrated Circuit (IC) by comparing it with trusted or reference design has been investigated extensively for almost a decade \cite{xiao2016hardware}. Unfortunately, these post-silicon detection techniques do not apply to COTS components due to the following reasons:


\begin{itemize}
    \item In the COTS components, golden references are not available whereas they can be obtained in the case of the custom ICs.
    \item When inserting Trojans into COTS components, adversaries have more liberty in altering the power, area, and timing constraints as these parameters are set by the designers themselves; allowing an adversary to insert more complex and large Trojans in COTS ICs.
\end{itemize}


Static and dynamic analysis techniques that apply to third-party IPs (3PIPs) do not apply to COTS ICs for the following reasons:
\begin{itemize}
    \item Usually, static analysis requires white box access to the netlist to perform analysis of the structural and functional features of the internal components \cite{fanci}, \cite{hoque2018hardware}. However, in the case of COTS ICs, rigorous ICs reverse engineering is needed to obtain netlists. 
    \item For dynamic approaches like logic testing, gate-level netlists are required to generate effective test vectors that activate the Trojan. Therefore, such dynamic techniques are inapplicable as well. 
\end{itemize}


Due to these challenges, there is a dearth of countermeasures in addressing the trust issues in COTS components \cite{xiao2016hardware}. Since none of the entities such as the design house, foundry, or testing facility can not be trusted, the supply chain of the COTS component represents a zero-trust threat model. Department of Defense (DoD) has recently emphasized the need for zero trust assumptions while procuring microelectronics \cite{collier2021zero}. Therefore, novel research is needed to enable trustworthy system design using COTS IC under zero trust assumptions.  

Confidentiality attacks through hardware Trojans within a processor core could leak secret information such as cryptographic keys  from internal registers and data path signals. These attacks are particularly channeling to detect as leakage could be enabled without corrupting the correct functionality of the system~\cite{lin2009moles}. In this paper, we propose $HOACS$ (\textbf{H}omomorphic \textbf{O}bfuscation \textbf{A}ssisted \textbf{C}oncealing of \textbf{S}ecrets), a Residue Number Coding (RNC)-based software transformation framework that enables the execution of the program while the data stays in encoded form. Such encoded execution of data can protect the secrets (e.g., encryption key, private data, password) from being leaked by hardware Trojans inside the untrusted microprocessor. 
RNC was originally coming from a residue number system (RNS), a non-weighted number system that represents numbers using the remainders obtained when dividing by a set of relatively prime numbers. Furthermore, it was developed to allow the implementation of arithmetic hardware operations without requiring extensive carry operations. Most arithmetic operations are easily supported by an RNC system.
RNC supports homomorphic operations for multiplication, addition, subtraction, and all the operations that can be performed by leveraging them (e.g., left shift of $n$ positions can be performed through a multiplication by $2n$) \cite{7174809_RNC_obfus}. All other operations require decoding of the data before operation and can be encoded back. 
 $HOACS$ transforms the program to be executed on untrusted microprocessors such that only the RNC encoded version of the sensitive data is used during a desired period of execution. This encoded data is unintelligible unless decoded through a series of operations. The sensitive data can be decoded at a stage defined by the user during the transformation of the program. During this protected execution period, hardware Trojans that snoop sensitive data and leak them through functional paths (e.g., memory bus, output ports) or side-channel (e.g., MOLES Trojan \cite{lin2009moles}) can only get access to assets in an unintelligible form. Although the author in Ref. \cite{LRA} has discussed the RNS property's resistance to side-channel data leakage, to the best of our knowledge, this is the first instance of introducing a software-based framework designed to protect against hardware Trojan attacks. Although data protection using homomorphic encryption is a commonly used technique, deploying it on resource-constrained embedded hardware remains a challenge. However, our proposed RNC-based framework supports homomorphic properties to protect data with lower overhead, offering a solution for safeguarding data in COTS processors of embedded systems. In this paper, we present the following major contributions:
\begin{enumerate}
    \item To the best of our knowledge, this is the first a software-based solution designed to prevent hardware Trojans from leaking sensitive data, taking into account a zero-trust supply chain for COTS processors. Unlike most existing countermeasures, HOACS does not require any analysis or modification of the IC design and operates under the assumption that all major IC supply chain entities, including design, fabrication, and testing facilities are untrusted. 
    \item We have implemented the $HOACS$ framework that applies the required code transformations for enabling RNC encoded execution for C implementation of Advanced Enc Standard (AES) cryptographic program. 
   \item To further automate $HOACS$ and reduce its overhead, we have leveraged the LLVM compiler tool-chain to develop $HOACS-IR$, an LLVM plugin that applies the required code transformations at the intermediate representation of AES and can be extended for other programs. 
    
    \item To evaluate the security against leakage Trojans, we have executed the $HOACS$ transformed AES on an x86 processor using the gem5 simulator for different values of the asset. By performing an extensive search in internal registers of the processor, we confirmed the confidentiality of the asset against Trojan attacks as any of the raw key does not show up in the registers.
    \item Additionally, We have also analyzed the feasibility of the burte-force attacks and hardware Trojan attack. If an attacker collects all the encoded data and attempts to apply a brute-force attack to decode it, our analysis suggests that it would take an impractically long amount of time to brute-force decode the sensitive data protected by our framework. 
    
    \item Finally, we compared the performance of $HOACS$ with other state-of-the-art fully and partially homomorphic methods. Results suggest that $HOACS$ enables significantly faster execution of the major operations, making it a viable solution for COTS processors.
\end{enumerate}

\par

The rest of the paper is organized as follows: Section II provides the threat model, background on existing hardware Trojan countermeasures, and RNC systems. Section III describes the proposed framework, an implementation case study on AES, and LLVM-based automation. Finally, in Section IV, we present a detailed security analysis along with a comparative study on the computational performance and conclude in Section V. 

\vspace{-0.3cm}

\section{Background}
\subsection{Threat Model}
\begin{figure}[t!]
\centering \includegraphics[width=0.95\linewidth]{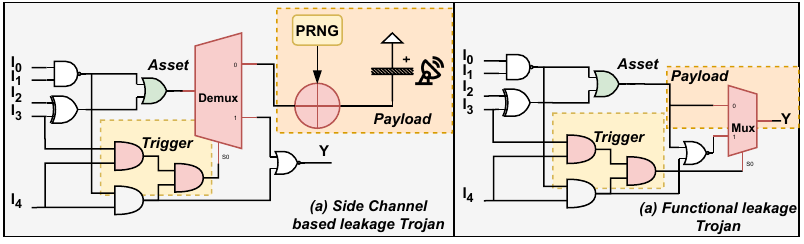}
    \caption{
     Hardware Trojan examples that leak sensitive information through (a) side channel and (b) functional port $Y$ observable to an attacker. The leakage activates when the trigger condition is applied to the design.   
}
\label{fig:leakage_trojan}
\end{figure}

The HOACS framework provides software-based security to the COTS ICs, which have been designed and manufactured in an untrusted environment. The following scenarios have been considered within this framework.
\begin{itemize}
    \item The threat model considers all supply chain entities, starting from the design house to the testing and packaging facility in Fig. \ref{fig:ch1_customIC} to be untrusted.
    \item The entities procuring the untrusted COTS IC and integrating the software code and the corresponding compiler toolchain are considered trusted. That means the attacker does not have any information about the encoding parameters, such as RNC moduli.
    \item The HOACS framework has been designed to protect against data leakage within the processor's registers, assuming that the attacker aims only to leak sensitive data. While this should also protect secret data passing through data path signals, we have not evaluated this aspect in this paper. 
    \item The leakage path could be functional (e.g., output ports) or side channel (power consumption) that is triggered during the execution of specific program which can be used for secret asset leakage. An example of these two classes of leakage Trojan is shown in Fig. \ref{fig:leakage_trojan}. 
\end{itemize}



\subsection{Existing Hardware Trojan Countermeasures}

Existing hardware Trojan countermeasures are developed to address one of the two major threat models: i) untrusted foundry and ii) untrusted IP vendors. To detect Trojans inserted by untrusted foundry, researchers have followed strategies that rely on side-channel analysis  \cite{hoque2017golden, rad2008sensitivity, jin2008hardware, soll2014based, forte2013temperature} or logic testing\cite{chakraborty2009mero, banga2008region}, as illustrated in Table \ref{comp_existing_cntmsr}. 
The majority of these methods depend on one or more of the following conditions: A golden design, a golden side-channel signature, and design-time adjustments (to implant sensors or expand the number of observable/controllable spots). These requirements cannot be met in the COTS IC supply chain, making the techniques inapplicable for such components. The necessity for design-time change also nullifies methods based on split manufacturing \cite{vaidyanathan2014efficient, imeson2013securing}, hardware obfuscation \cite{chakraborty2009security, dupuis2014novel, yasin2015transforming}, and design-for-security \cite{cao2014cluster, jin2012post, xiao2015efficient} that are thought to be successful in avoiding Trojan insertion at the untrusted foundry.

\begin{table*}[t!]
\centering
\caption{Comparison of $HOACS$ with Existing Hardware Trojans Countermeasures.}
\label{comp_existing_cntmsr}
\resizebox{\textwidth}{!}{%
\begin{tabular}{c|c|c|c|c|c}
\hline
Approaches                             & Reference                                                                                   & Detection~/      & HT         & HT         & HT in      \\
                                       &                                                                                        & Tolerance Phase  & in IP      & in IC      & COTS IC    \\ \hline
\cellcolor{gray!75} Proposed Method   & $HOACS$                                                                     & Runtime          & \checkmark & \checkmark & \checkmark\\ \cline{2-6}
\cellcolor{gray!75} Data Obfuscation    & DETON \cite{cassano2022deton}                                                                    & Runtime          & \checkmark & \checkmark & \checkmark \\ \cline{2-6} 
\cellcolor{gray!75} Software Variant   & TReC\cite{hasan2022Trojan}                                                                    & Runtime          & \checkmark & \checkmark & \checkmark \\ \cline{2-6} 
\cellcolor{gray!75} IC Redundancy      & SAFER PATH \cite{ref:safer}                                                            & Runtime          & \xmark     & \checkmark & \checkmark \\ \cline{2-6} 
\cellcolor{gray!75} IP Redundancy      & TrojanGuard \cite{malekpour2017Trojanguard}, HLS \cite{rajendran2016building}          & Runtime          & \checkmark & \xmark     & \xmark     \\
\cellcolor{gray!75}                    & Task Scheduling \cite{ref:mpsoc}, \cite{wang2018security}                              &                  &            &            &            \\ \hline
\cellcolor{gray!25} Static Analysis    & ANGEL \cite{fyrbiak2018hal}, COTD \cite{ref:COTD}                                           & Pre-Silicon      & \checkmark & \xmark     & \xmark     \\  \cline{2-6} 
\cellcolor{gray!25} Formal Methods     & Proof Check \cite{love2012proof}, Verification \cite{rajendran2015detecting}           & Pre-Silicon      & \checkmark & \xmark     & \xmark     \\ \cline{2-6} 
\cellcolor{gray!25} IP Monitoring      & Many-core \cite{kulkarni2016svm}, ISA Power \cite{lodhi2017power}                      & Runtime          & \checkmark & \xmark     & \xmark     \\ \cline{2-6} 
\cellcolor{gray!25} Side-Channel       & Power \cite{hoque2017golden}, \cite{rad2008sensitivity}, Delay \cite{jin2008hardware}, & Post-Silicon,    & \xmark     & \checkmark & \xmark     \\ \cline{2-6} 
\cellcolor{gray!25} Logic Testing      & MERO \cite{chakraborty2009mero}, Region Based \cite{banga2008region}                   & Pre/Post-Silicon & \xmark     & \checkmark & \xmark     \\ \hline
\cellcolor{gray!5} Design-for-Sec.     & Sensors \cite{cao2014cluster}, \cite{jin2012post}, OBISA \cite{xiao2015efficient}      & Pre/Post-Silicon & \xmark     & \checkmark & \xmark     \\ \cline{2-6} 
\cellcolor{gray!5} Hardware  obfuscation          & Logic Encryption \cite{dupuis2014novel},                                               & Pre-Silicon      & \xmark     & \checkmark & \xmark     \\ 
\cellcolor{gray!5}     &  HARPOON \cite{chakraborty2009security}, Camouflaging \cite{yasin2015transforming}       &   &   &   &     \\  \cline{2-6}
\cellcolor{gray!5} Split Manufacturing & \cite{vaidyanathan2014efficient}, 3D integration \cite{imeson2013securing}             & Pre-Silicon      & \xmark     & \checkmark & \xmark    \\ \hline
\multicolumn{5}{l}{Class of Countermeasure:
                 ~~\tikz\draw[black,fill=gray!85] (0,0) circle (.8ex); Trojan Tolerance,
                 ~~\tikz\draw[black,fill= gray!45] (0,0) circle (.8ex); Trojan Detection,
                 ~~\tikz\draw[black,fill= gray!10] (0,0) circle (.8ex); Trojan Prevention

}\\ 
\end{tabular}%
}
\end{table*}

On the other hand, to enable verification of Trojans inserted by untrusted IP vendors, several methods have recently been developed to identify or mitigate Trojans implanted in soft IPs (RTL/gate-level netlist). However, white-box accessibility to the hardware IP is necessary for detection-oriented solutions like static code analysis\cite{ref:fanci, ref:COTD, banga2010trusted, zhang2015veritrust} and formal methodologies\cite{love2012proof, rajendran2015detecting}. Redundancy-based approaches \cite{malekpour2017Trojanguard,rajendran2016building} and runtime Trojan monitoring \cite{kulkarni2016svm, lodhi2017power} can be integrated to enable runtime tolerance and detection in untrusted IPs. However, these techniques either require IP vendor diversity and/or modification of the hardware code. Such methods are impractical for COTS components due to the  need for design-time modification and increased cost to enable redundancy. 

To address the trust issue in COTS processors, researchers have used hardware redundancy-based techniques that enable Trojan-tolerant computing in the field. For instance, SAFER PATH \cite{ref:safer} simultaneously executes the program in multiple processing elements and compares the results to detect/tolerate Trojan activation through majority voting. However, such techniques go against the economic motivation of using COTS hardware. 
 Mavroudish et al. have proposed a high-assurance architecture, known as “Myst”, combining protective redundancy with modern threshold cryptographic techniques \cite{mavroudis2017touch}. Similar to SAFER PATH, Myst relies on hardware redundancy and vendor diversity. Software-based solutions have also been proposed to bypass Trojan activation \cite{marcelli2018defeating}. However, such techniques may not be able to bypass Trojans that activate based on data instead of the instruction sequences. In \cite{hasan2022Trojan} a software-redundancy based technique has been proposed to enable Trojan resilience in COTS processors. This technique targets Trojans that impact the functionality by enabling multi-variant execution of the same program, followed by majority voting of the output of the variants. However, these two software-based techniques do not protect against information leakage. To address the issue of such leakage Trojans in COTS hardware, Meng et al. proposed a pre-deployment solution that investigated the secret asset flow for custom ICs that are connected to COTS ICs in the system by analyzing the custom IC designs \cite{meng_cots_IF}.

To defeat Trojans in COTS hardware that leak information, \cite{ref:COTS} incorporates fragmented program execution across selected PEs where each unit executes a portion of the code. Such context switching limits the access of individual PEs to critical data for specific applications. However, it is difficult to guarantee that it can ultimately prevent side-channel leakage of any asset processed within an arbitrary program.  Another proposed solution~\cite{cassano2022deton} obfuscates the software code to enlarge the set of registers used during program execution and spread the sensitive information through several registers and instruction cycles. The assumption is that such obfuscation will cause information-stealing Trojans to monitor a large set of registers and to monitor them for a long time to obtain confidential data. $HOACS$ provides more robust protection of secret data compared to DETON~\cite{cassano2022deton} by completely eliminating the confidential data during the target execution period through mathematically robust homomorphic transformation instead of distributing it in space in time. Moreover, unlike \cite{ref:COTS, ref:safer, mavroudis2017touch}, $HOACS$ does not rely on hardware redundancy and vendor diversity.

\subsection{Residue Number Coding (RNC)}
\label{RNC}
\subsubsection{\textbf{Basic of RNC}}

In RNC system, a number $v$ can be splitted into its residue numbers $v_i$ with respect to $u$ number of pairwise co-prime moduli $m_0 < m_1 < m_2 ...... <m_{u-1}$. The $u$ number of the residue of $v$ can also be considered as $u$-digit number. For the $i_{th}$ number of residue, the notational representation shown in Equation \ref{RNC_equation} \cite{RNC_BOOK}.
\begin{equation}
v_i=v \bmod m_i=\langle v\rangle_{m_i}
\label{RNC_equation}
\end{equation}
An RNC represented number should be greater than the product $M$ of $u$ co-prime moduli numbers. Here, $M= m_0 \times m_1 \times \cdots \times m_{u-2} \times m_{u-1}$, which is also known as the dynamic range. For instance, if $u=3$ and $m_1=12,$ $m_2=14$ $and$ $m_3=7$ the dynamic range will be $M= 12 \times 14 \times 7 = 1176$. Also, the range of residue represented number will be $v \in 0, 1, 2, 3, \cdots , 1175$. Using larger $m$-values, or more of them, allows a larger range of values to be represented. Although it is simplest to encode unsigned integers, signed integers also can be easily represented by adding an offset to slide the range of numbers to start at zero. In the decoding process, this shift would have to be reversed. For the signed integer the mathematical notation can be represented by Equation \ref{signed_rnc_equation}  \\
\begin{equation}
\langle-v\rangle_{m_i}=\langle M-v\rangle_{m_i}
\label{signed_rnc_equation}
\end{equation}

Based on the inherent nature of residue number systems, the encoded result of any number smaller than either $m$-value will include at least one instance of its original value. For example, the result of encoding the value 5 using modulus 4 and 7 has components 1 and 5.
Modular arithmetic allows for random multiples of each $m$-value to be added to its corresponding residue component without changing its effective value \cite{7174809_RNC_obfus}. This property can be used to provide increased protection in data obfuscation systems. Selected $m$-values must be pairwise co-prime for the transformation to be reversible. 

\begin{figure}[!t]
    \centering   \includegraphics[width=0.9\linewidth]{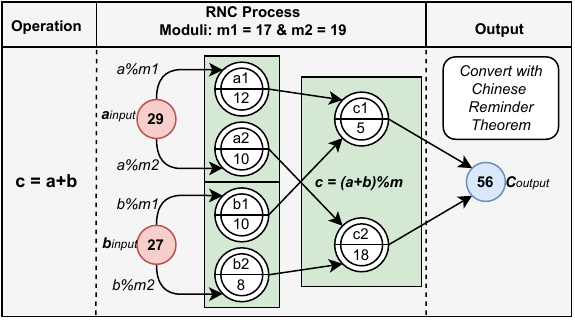}
        \caption[Example of Residue Addition]{
        A simple RNC process using an example similar to that used 
        by Garner in his seminal paper \cite{5219515_Garner_ResidueNumberSystem}.  
        This example demonstrates the homomorphic addition of integers 29 and 27 using moduli 17 and 19.  
    }
    \label{fig:rncAdd}
\end{figure}

\subsubsection{\textbf{Example of Arithmetic Operation in RNC}}
To provide an example of RNC arithmetic operations, we have demonstrated a step-wise illustration of the addition operation in Fig. \ref{fig:rncAdd}. In this example, two decimal numbers have been taken to obfuscate using RNC encoding, and after RNC addition, the numbers have decoded again to their decimal representation. The moduli $m_1=17$ and $m_2=19$ has been taken for this example. According to RNC conversion, the remainder of the given number with respect to moduli results in the RNC encoded representation. Hence, for 29 and 27 the RNC conversion is as follow: $(12 \equiv 29 \bmod 17)$ and $(10 \equiv 29 \bmod 19)$, $(10 \equiv 27 \bmod 17)$ and $(8 \equiv 27 \bmod 19)$. The arithmetic addition operation of these two RNC encoded numbers is the following. 

$$
\begin{aligned}
29 & \equiv(12 \mid 10)_{\mathrm{RNC}} \\
+27 & \equiv(10 \mid 8)_{\mathrm{RNC}} \\
\hline 56 & \equiv(5 \mid 18)_{\mathrm{RNC}}
\end{aligned}
$$

The RNC addition operation produces the results in encoded form, and to decode these results, the Chinese Remainder Theorem (CRT) has been adopted~\cite{rnsTutorialReview}. Equation \ref{CRT} represents the CRT-based decoding that has been followed in our proposed framework.
\begin{equation}
v=\left(v_{u-1}|\cdots| v_2\left|v_1\right| v_0\right)_{\mathrm{RNC}}=\left\langle\sum_{i=0}^{u-1} M_i\left\langle\alpha_i u_i\right\rangle_{m_i}\right\rangle_M
\label{CRT}
\end{equation}

Here, $M_i=M / m_i$, and $\alpha_i=\left\langle M_i^{-1}\right\rangle_{m_i}$ \vspace{0.2cm}is the multiplicative inverse of $M_i$ with respect to $m_i$ \cite{RNC_BOOK}. CRT-based decode methods operate on each component separately until the final stages, when the components are combined to produce a decoded value. The weighted mixed-radix conversion (MRC) process has been used to obtain positional weights. Since each residue component is independent, these methods must add positional weighting. As shown in Algorithm \ref{algo_flow}, the extended Greatest Common Divisor (GCD) is needed as part of the MRC process.  Specifically, the first coefficient is needed to add positional significance to the RNC components.

\subsubsection{\textbf{Homomorphic Encryption using RNC}}

\begin{figure}[!t]
    \centering
    \includegraphics[width=.95 \linewidth]{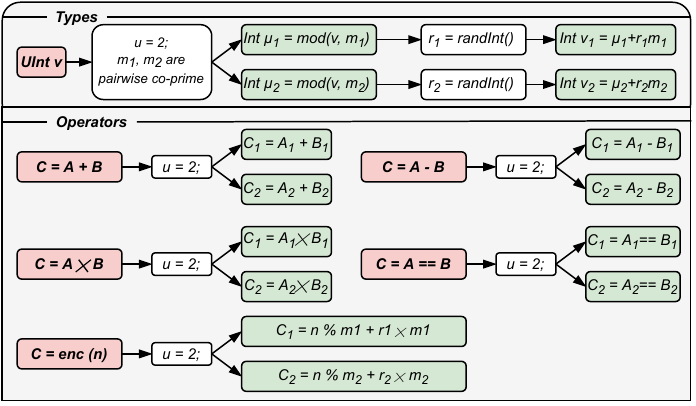}
        \caption[Example of Residue Addition]{This figure visualizes several RNC conversions.
                Red nodes represent types of operations that require RNC processing.
                The green nodes represent operations that have been fully translated to use residue-based values.  
    }
    \label{fig:rnc_operations}
\end{figure}

The aforementioned example has demonstrated that the given data are obfuscated through the RNC system with respect to user-defined moduli and the number of pairwise co-primes. These values are essential for decoding the obfuscated data. Consequently, even if an attacker were to gain access to the encoded data, deciphering it would remain unattainable. Figure \ref{fig:rnc_operations} depicts the fundamental arithmetic operations within the RNC framework. Although addition, subtraction, multiplication, comparison, and encoding are illustrated in Fig. \ref{fig:rnc_operations}, the proposed \textit{HOACS-C} library also encompasses additional logical and arithmetic operations, as detailed in Table \ref{table-data-obf-cap}. This substantiates the claim that RNC operations exhibit fully homomorphic characteristics. Fully Homomorphic Encryption (FHE) constitutes an encryption scheme that facilitates the execution of computations directly on encrypted data without necessitating prior decryption. A fully homomorphic encryption scheme permits arbitrary computations on ciphertexts, thereby supporting addition and multiplication operations on encrypted data. These dual operations form the foundation for a broad spectrum of complex computations and can amalgamate to execute any function that can be represented as a polynomial \cite{HORNS}. The RNC-based operations under discussion support both addition and multiplication, as well as logical and exponential operations~\cite{HORNS}. As a result, the RNC encoding and RNC-based homomorphic execution render the proposed concept increasingly appealing since data confidentiality can be ensured without modifying or assessing the hardware. Due to the inherent properties of Residue Numbers, the complexity of individual homomorphic arithmetic or logical instructions exhibits significant variation.

\section{Proposed Method}
    \subsection{Overview of Framework} 

The \textit{HOACS} framework facilitates secure program execution through a series of well-defined stages, as illustrated in Figure 5. This overview provides clarity on each step within the framework:

\paragraph{\textbf{Initializations}} The framework begins by establishing essential parameters. It identifies the secret assets that require encryption and defines the co-prime moduli, which are user-defined. This flexibility allows users to specify both the quantity and the values of the moduli. For example, Algorithm 1 highlights the initial setup where sensitive variables (secret assets) and m-values (moduli) are defined.

\paragraph{\textbf{Encoding Secret Assets}} In this phase, sensitive variables undergo transformation into RNC-based data. This transformation involves modular arithmetic operations, as depicted by the \textit{ENCODE} function in Algorithm 1. This function demonstrates how it encodes the input data, producing an RNC-based version of the sensitive information.

\paragraph{\textbf{Secure Program Execution}} With the data now in RNC format, the framework executes the program, accommodating both arithmetic and logical operations. This step does homomorphic execution as data used in this step are encoded. The ADDENC function in Algorithm 1 exemplifies an addition operation within the HOACS framework. This example underlines the framework's capability to adapt any C program to its RNC-based equivalent. Notably, during this execution phase, any hardware Trojan attempting to access the processor's internal registers and wires would encounter only encoded values, rendering the data unintelligible to attackers. Depending on the application or threat model, protection might be necessary only for specific program parts, such as during key expansion in the AES algorithm discussed later. The framework's adaptability is further supported by the inclusion of common arithmetic and Boolean RNC algorithms in the appendix.

\begin{figure}[t!]
\centering \includegraphics[width=0.95\linewidth]{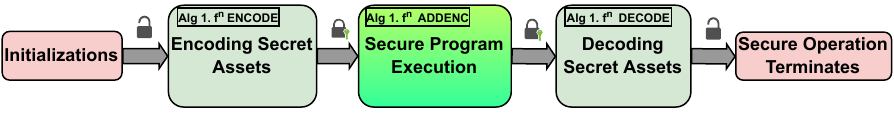}
    \caption[$HOACS-C$ Design Ideas]{
    Proposed framework to obfuscate secret assets. Also, each stage shows its functions from Alg \ref{algo_flow}. $ADDENC$ function is one of the arithmetic operations shown here in the execution stage.  
}
\label{fig:rnc_framework}
\end{figure}

\paragraph{\textbf{Decoding Secret Assets}} After securely executing the transformed data, the decoding process employs the Chinese Remainder Theorem, as detailed by the DECODE function in Algorithm 1. This step reverses the encoding to retrieve the original data from its encoded state. After this step all subsequent operations will be non-homomorphic.

\paragraph{\textbf{Termination of Secure Operation}} The final stage marks the end of secure operations, with a thorough cleanup of all used moduli from memory to maintain security and prevent any data residue. 

This systematic approach ensures the \textit{HOACS} framework provides effective protection for sensitive data during program execution. By leveraging the RNC, it offers a robust defense against Trojan attacks, optimizing for both security and efficiency. We have also developed an LLVM compiler extension, \textit{HOACS-IR}, enables automatic replacement of operations with RNC equivalents at the intermediate representation level, requiring minimal semantic modifications to source code. This approach improves performance and allows easier integration of RNC-based protection by using variable name prefixes to identify sensitive data. The proposed framework considers the following assumption in terms of securing assets:
\begin{enumerate}
    \item The proposed framework, as presented in this paper, ensures secure execution once the assets are encoded using moduli $(m_1, \cdots, m_n)$.
    \item It is postulated that the Trojans will be activated during the execution of security-critical operations on the processor.
\end{enumerate}
The rationale behind these assumptions is that processors handle numerous operations simultaneously, many of which do not hold significant value for potential adversaries. Attackers are primarily interested in capturing confidential information utilized in security-critical operations. Even if the attacker manages to activate the Trojan prior to the execution of these critical operations, the efficient identification of sensitive data remains challenging due to the vast amount of data chunks being processed by the registers from various programs.
To solve this problem, our proposed framework, \textit{``HOACS-IR"}. This framework can be able to encode sensitive assets during the compilation process, thereby ensuring that the program operates in an encrypted mode consistently while the system is in operation. Below we describe detailed implementation for a program written in C, case study implementation for AES, and automation through LLVM.    

    \subsection{Implementation} 

\begin{algorithm}
\caption{$HOACS-C$ Execution Flow}\label{algo_flow}
\begin{algorithmic}[1]
\Function{main}{int argc, char *argv[]}
\State $\text{unsigned int a, b, c}$ \Comment{Sensitive Variables}
\State $\text{unsigned int m1, m2}$ \Comment{m-values}
\State $\textit{Encoded aEnc} \gets \textit{\textbf{encode(a, m1, m2)}}$
\State $\textit{Encoded bEnc} \gets \textit{\textbf{encode(b, m1, m2)}}$
\State $\textit{Encoded cEnc} \gets \textit{\textbf{AddEnc(aEnc, bEnc, m1, m2)}}$
\State $\textit{int cDec} \gets \textit{\textbf{decode(cEnc, m1, m2, NULL, NULL)}}$

\Comment{Decoded Data Can Be Used As Needed}
\State \Return 0
\EndFunction

\Function{encode}{v, m1, m2}:
\State $\textit{Encoded} \gets \textit{e}$
\State $\textit{e.u1} \gets \textit{v \% m1}$
\State $\textit{e.u2} \gets \textit{v \% m2}$
\State $\textit{e} \gets \textit{\textbf{addRand(e, m1, m2)}}$
\State \Return e
\EndFunction

\Function{AddEnc}{x, y, m1, m2}
\State $\textit{    Encoded} \gets \textit{z}$
\State $\textit{    z.u1} \gets \textit{(x.u1 \% m1) + (y.u1 \% m1)}$
\State $\textit{    z.u2} \gets \textit{(x.u2 \% m1) + (y.u2 \% m2)}$
\State $\textit{    z} \gets \textit{\textbf{addRand(z, m1, m2)}}$
\State \Return z
\EndFunction

\Function{decode}{Encoded e, m1, m2, int* v1, int* v2}
\State $\textit{int M} \gets \textit{m1} \times \textit{m2}$
\State $\text{int x, y}$
\State $\textbf{extended\_gcd(m1, m2, \&x, \&y)}$
\State $\textit{int loc\_v1} \gets \textit{(e.u1) \% m1}$
\State $\textit{*v1} \gets \textit{loc\_v1}$
\State $\textit{int loc\_v2} \gets \textit{(e.u2 - loc\_v1)} \times \textit{x}$

\While{$loc\_v2 < 0$}
    \State $loc\_v2 \gets \textit{loc\_v2  +  m2}$
\EndWhile
\State $\textbf{end while}$

\State $\textit{*v2} \gets \textit{loc\_v2}$
\State $\textit{int v} \gets \textit{loc\_v1 + loc\_v2} \times \textit{m1}$
\State $\textit{v} \gets \textit{v \% M}$
\State \Return v
\EndFunction

\Function{extended\_gcd}{a, b, *x, *y}
\If{$a == 0$}
    \State $*x \gets 0$
    \State $*y \gets 1$
    \State \Return b
\EndIf
\State $\textbf{end if}$

\State $\text{int loc\_x, loc\_y}$
\State $gcd \gets \textit{\textbf{extended\_gcd(b \% a, a, \&loc\_x, \&loc\_y)}}$
\State $*x \gets loc\_y - \frac{b}{a}  \times loc\_x$
\State $*y \gets loc\_x$
\State \Return gcd
  
\EndFunction

\Function{addRand}{Encoded X, m1, m2}
\State $\textit{    X.u1} \gets \textit{\textbf{rand()} * m1}$ 
\State $\textit{    X.u2} \gets \textit{\textbf{rand()} * m2}$
\State \Return X
\EndFunction

\end{algorithmic}
\end{algorithm}

To implement the HOACS, we introduced a library tailored for the C programming language, named \textit{HOACS-C}. This framework enables the utilization of arithmetic and logical operations, as detailed in Table~\ref{table-data-obf-cap}, and supports the use of composite data types, such as \texttt{structs}, alongside primitive data types. The selection of the C language was motivated by its extensive adoption in embedded systems and its direct interaction with hardware components.

The procedural execution flow of \textit{HOACS-C} is meticulously described in Algorithm \ref{algo_flow}, referencing a specific example discussed in Section \ref{RNC}. To effectively apply the \textit{HOACS-C} framework, users are guided through the following steps:

\begin{enumerate}
    \item \textbf{Library Inclusion}: Initiate by incorporating the \textit{HOACS-C} library into the project. This step provides access to a comprehensive suite of homomorphic encryption functions and data structures essential for secure data operations.
    
    \item \textbf{Initialization of RNC Component}: Configure the desired moduli (\texttt{m1} and \texttt{m2}) values. This crucial step allows for the customization of the security level and operational efficiency, aligning with the specific needs of the application. For illustration, in the main function of Algorithm 1, two unsigned integers are added and the result is stored. Despite \textit{HOACS-C}'s support for variable sizes through pre-processor definitions, our emphasis has been on 8-bit unsigned integers to minimize m-values for testing. While only two moduli (\texttt{m1} and \texttt{m2}) are used for simplicity, the framework theoretically supports any number, with an increase in the quantity of moduli correlating to enhanced security. The detailed analysis of this security feature is shown in Section \ref{RNC_SEC}.
    \item \textbf{Selecting the Secret Data}: In this step, users must select the data variables they wish to protect from hardware Trojan attacks. Users can choose any number of variables for protection. As illustrated in Algorithm~\ref{algo_flow}, we have chosen the unsigned integers \texttt{a}, \texttt{b}, and \texttt{c} as our secret data. In a later section of this paper, during our security analysis, we treat the plain key of the AES algorithm as our secret data. When this plain key is passed to the AES \texttt{key expansion} function, it is encoded for protection.
    \item \textbf{RNC Implementation Selection}: The \textit{HOACS-C} framework offers three variants of RNC implementation, enabling the creation of functionally equivalent programs capable of homomorphic operations:
    \begin{itemize}
        \item \textbf{RNC Component}: Provides the foundational functionality for RNC encoding, the execution of individual homomorphic RNC operations, and the decoding of RNC data back to its original type.
        \item \textbf{RNC-Tree Component}: Offers a binary search tree implementation utilizing RNC encoded values, employing the RNC Less-Than operator to shift lookup complexity from linear to logarithmic.
        \item \textbf{RNC-Grid Component}: Introduces a data structure with constant time lookup at the expense of additional memory space overhead, utilizing RNC components for indexing into a multi-dimensional array based on the selected \texttt{m}-values.
    \end{itemize}
    
    \item \textbf{Data Encoding}: Utilize the encode function to transform plain data into its RNC-encoded form, readying it for secure processing. As illustrated in Fig.~\ref{fig:rncAdd}, encoding the data variables \texttt{a} and \texttt{b} with RNC yields four obfuscated variables (\texttt{a1}, \texttt{a2}, \texttt{b1}, \texttt{b2}). These variables are then subjected to arithmetic and logical operations. In the event that an attacker manages to probe the registers and obtain these encoded variables, decoding them would be infeasible without knowledge of the moduli employed in the encoding process. Fig.~\ref{fig:rnc_operations} additionally details the operational flow of other arithmetic and logical operations supported by \textit{HOACS}.

    \item \textbf{Secure Operations Execution}: Conduct operations on the encoded data, such as \texttt{AddEnc}, for secure arithmetic or logical processing shown in Algorithm \ref{algo_flow}. This step's versatility is demonstrated through the application of the \textit{HOACS-C} library to various programs, including an in-depth analysis of the AES encryption algorithm's KeyExpansion function as a case study for secure computation.
    
    \item \textbf{Decoding of Results}: Utilize the decode function post-computation to revert RNC-encoded data back to its original format, rendering the results accessible and interpretable.
\end{enumerate}

This structured methodology bolsters data security and ensures operational efficiency. By leveraging \textit{HOACS-C}, developers are equipped to implement a flexible and robust system for embedding homomorphic operations, significantly benefiting applications that necessitate data protection in resource-constrained environments.

\begin{table*}[t]
\renewcommand{\arraystretch}{1.5}
\setlength{\tabcolsep}{1.0\tabcolsep}
\centering
\caption{RNC-based arithmetic and logical operations implemented in $HOACS-C$  and data obfuscation processes}
\begin{tabular}{|p{0.15\linewidth}|p{0.20\linewidth}|p{0.55\linewidth}|} 
    \hline
    Operation           & Homomorphism         & Data Obfuscation Process                             \\    
    \hline \hline
    Addition            & Fully Homomorphic     &    (1) Remove random multiple shifting, (2) add RNC components (3) check overflow \\    
    \hline
    Subtraction         & Fully Homomorphic     & (1) Remove random multiple shifting, 
                                                    (2) subtract RNC components, 
                                                    (3) check for underflow                                 \\    
    \hline
    Multiplication      & Fully Homomorphic     & (1) Remove random multiple shifting, 
                                                    (2) multiply RNC components, 
                                                    (3) check for overflow                                  \\    
    \hline
    Less-than Operator  & Partially Homomorphic & (1) partially decode to get the mixed-radix terms 
                                                (2) compare corresponding positions of mixed-radix values 
                                                \cite{knuth97_ACP_vol2}   \\    
    \hline
    Division            & Partially Homomorphic & (1) iteratively subtract divisor from numerator (2) return no. of subtraction cycles needed to make numerator smaller than divisor; Note: this relies on the partially homomorphic `less-than' operator                                   \\    
    \hline
    Modulus             & Partially Homomorphic & (1) iteratively subtract divisor from the numerator 
                                                (2) return the numerator when it is smaller than the divisor; Note: this relies on a partially homomorphic `less-than' operator
                                                             \\    
    \hline
    Exponential         & Fully Homomorphic     & (1) requires recursive calls of more elemental 
                                                homomorphic functions as listed in     
                                                Fig. \ref{fig:rnc_operations}                    \\    
    \hline
    Bit-wise XOR        & Partially Homomorphic & (1) expands into positional operations via Division 
                                                (2) results are combined                                    \\    
    \hline
    Equality Operator   & Fully Homomorphic     & (1) remove random multiple shifting 
                                                (2) compare RNC components                                  \\    
    \hline
\end{tabular}
\label{table-data-obf-cap}
\end{table*}

\subsection{Example Case Study: AES Key Expansion}
\begin{figure}[!t]
    \centering 
    \includegraphics[width=0.95\linewidth]{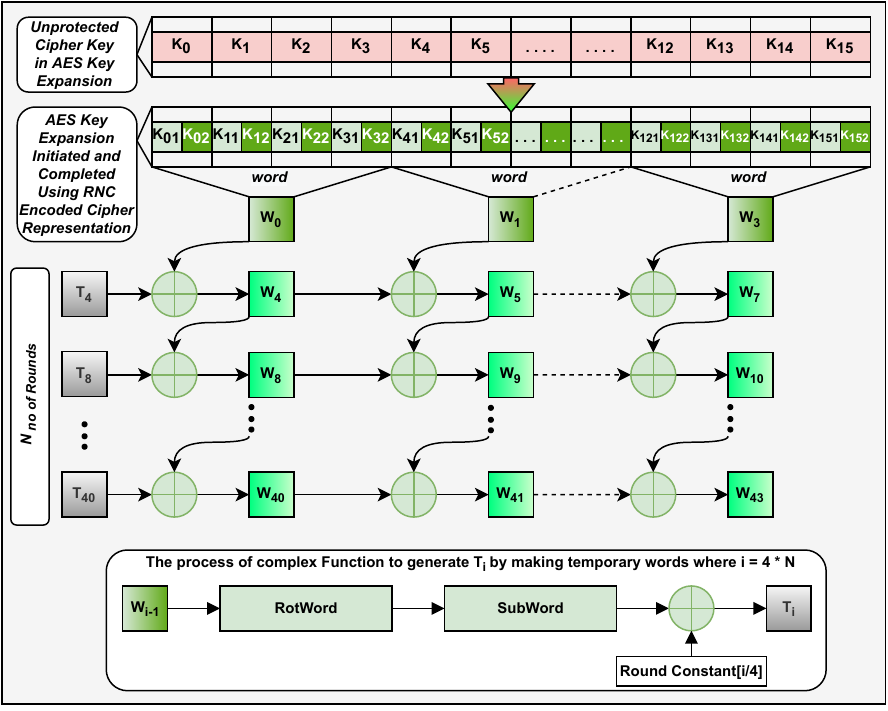}
    \caption[RNC Encoded AES Key Expansion]{
        AES key expansion after applying the proposed RNC obfuscation method. The RNC components are depicted by the two colors in each block of the key bytes ranging from $K_{01}$ to $K_{152}$. These RNC structs were used to store the initial key and further generated round keys throughout 
        the key expansion process. 
    }
    \label{fig:aesEncoded}
\end{figure}
An example is needed to evaluate the selected protection scheme's effectiveness. For this purpose, the key expansion phase of an AES implementation was selected for modification.  There were several reasons for this choice of the case study.  AES key expansion exemplifies a scenario for which an approach like $HOACS-C$ is well suited.  As the operations used with AES are defined at the byte level, the number of possible values, and therefore the $m$-values needed, is bounded to the range $0$ to $255$.  Although this restriction made implementing RNC operators easier, it was not required. Moreover, in the AES algorithm, key expansion is used to expand the plain key as shown in Fig. \ref{fig:aesEncoded}. This stage uses the input key to generate a series of successive round keys. The first round key is just the original key, and each successive round key is derived from a previous round key. Part of AES Key Expansion involves replacing parts of the key with values taken from the corresponding location in the SBox array.
This step directly uses the plain key, which makes this step most vulnerable to secret key leakage. If the attacker probes the registers during the execution of the key expansion module, the attacker can easily get access to the plain key. 


A C-language-based AES implementation has been selected as the test program for this paper \cite{tinyAES}. 
To enable the protection, the key is encoded prior to its processing in the key expansion function. The next step is to change the types of sensitive variables, which are the initial and generated round keys. The encoded initial key is illustrated in the second row of Fig. \ref{fig:aesEncoded}, and the generated round keys are shown as green boxes in each of the following rows.
The next step was to replace operations with their RNC equivalents, as shown in Algorithm \ref{algo:rnc_keyExpansion}.

 \begin{algorithm}
        \caption{KeyExpansion}
        \textbf{Global: } $nK$ \Comment{Number of 32-bit words in a key} \\
        \textbf{Global: } $nR$ \Comment{Number of AES Rounds} \\
        \textbf{Input: } Encoded[KeyExpLen] RoundKey \\
        \textbf{Input: } Encoded[nK] Key \\
        \textbf{Input: } $m_1, \ m_2$ 
        \begin{algorithmic}[1]
        \For {$i = 0; i < $ nK; $ ++i$ }
            \State RoundKey's word[$i$] $\gets$ Key's word[$i$]
        \EndFor
        \For {$n = nK; n < 4 \times (nR +1); ++n$}
            \State Initialize RoundKey[$n \to n +3$]
            \If {$ n \ \% \ nK == 0 $}
                \State RotWord()
                \State SubWord()
                \State XorEnc() First byte of current word
            \EndIf
            \If {$ n \ \% \ nK == 4 $} \Comment{Only used with AES 256}
                \State SubWord()
            \EndIf
            \State XorEnc() all bytes of current word 
        \EndFor
        \end{algorithmic} 
        \label{algo:rnc_keyExpansion}
    \end{algorithm} 

\subsection{Automation of $HOACS$ using LLVM IR}
\subsubsection{Motivation}
An LLVM compiler extension $(HOACS-IR)$ has been developed to apply the code transformation at the intermediate representation (IR) of the program, instead of the source code. The primary focus of this extension is to demonstrate that operations could be automatically replaced with their RNC equivalents without needing extensive semantic modifications to a program's existing source code. Providing such capability in the form of a compiler plugin extends the potential base of users who can easily integrate RNC-based protection in their existing code. Users will use prefixes in their variable names to identify sensitive data to apply this functionality. Integration in low-level abstraction also improves performance, as evident in our overhead analysis. 

\subsubsection{Coverage}
\textit{HOACS-IR} recognizes the sensitive variables identified by the user and traces out all instructions that operate on them. Although $HOACS$  can execute most of the arithmetic and logical operations, some operations can not be executed (such as exponential functions) by the current version of the $HOACS$ framework. Hence, if any instruction of a given program is not supported by the current version, the tool ensures every argument is decoded for that operation to preserve the behavior of the unmodified program. Almost all frequently used basic operations are supported by the current $HOACS-IR$ version, including addition, subtraction, multiplication, and equality comparison (more on Table \ref{table-data-obf-cap}).  

\subsubsection{Usage}
To apply this compiler extension, users identify sensitive data as explained above and then invoke the $HOACS-IR$ plugin using LLVM's ``opt'' tool. The ``opt'' tool then inserts RNC protection as a pass during LLVM's optimization phase. Once this pass has been initiated, the compiler extension identifies sensitive data through an ``rnc\_" prefix in user-defined variable names. Then it propagates protection markings to all implicitly sensitive variables downstream. Data obfuscation will later be applied to all explicit and implicit variables as necessary to protect sensitive data.



    
\begin{figure}[!t]
    \centering  
    \includegraphics[width=0.8\columnwidth]{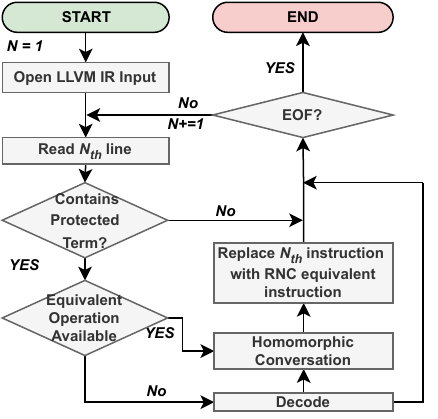}
    \caption{This flowchart shows the process of applying RNC transformations to the targeted LLVM IR program.}
    \label{fig:llvm}
\end{figure}


\subsubsection{Implementation}

Fig.~\ref{fig:llvm} illustrates the process of applying the \textit{HOACS} library using the LLVM IR extension. The step-wise explanation of this process is as follows:

\begin{enumerate}
    \item Automatically initialize sufficiently large \texttt{m}-values to encode 32-bit integers by iteratively checking pairs of values. Alternatively, users can define the \texttt{m}-values themselves.
    \item Iterate through all instructions in the LLVM IR input file.
    \item For each instruction, determine if any operands are marked as protected (with \texttt{`rnc\_'} prefix). Such marking indicates that the instruction requires RNC modification.
    \item If RNC modification is required, determine if an equivalent homomorphic RNC operation is available.
    \item If an equivalent operation is not available, any protected operands must be decoded. This preserves the semantics of the user's program for unexpected instructions.
    \item If an equivalent operation is available:
    \begin{enumerate}
        \item Ensure both operands are encoded. Any such operands not already identified as protected require encoding to enable homomorphic conversion of this instruction. This preserves the protection of operands/instructions already identified as sensitive.
            Even if neither operand is considered sensitive, both values must be encoded.
        \item Replace the existing LLVM IR instruction with the homomorphic RNC equivalent provided by this tool.
    \end{enumerate}
    \item Repeat these steps for all instructions.
\end{enumerate}

For RNC transformations to be reversible, the \texttt{m}-values used must be large enough to cover the entire range of possible values for the original unencoded data. To represent all unsigned 32-bit values, large \texttt{m}-values with a product (\texttt{M}) of at least $2^{32} - 1$ are required. This large \texttt{M} value necessitates each \texttt{m}-value being large as well. As the RNC encoding process is based on taking the modulus of the sensitive value by each \texttt{m}-value, any values smaller than either \texttt{m} value will not be obscured by the normal encoding process. To offset this issue, a "random multiple shift" is performed at compile-time for any constant values being encoded by the \textit{HOACS-IR} tool. For the \textit{HOACS-IR} implementation, \texttt{m}-values must be initialized before any RNC modifications can be made to the LLVM IR during the compilation process. This enables the compile-time encoding of static data, which has significant security benefits (brief discussion in Section \ref{HT_Sec}).

\begin{figure}[!t]
    \centering
    \includegraphics[width=0.95\linewidth]{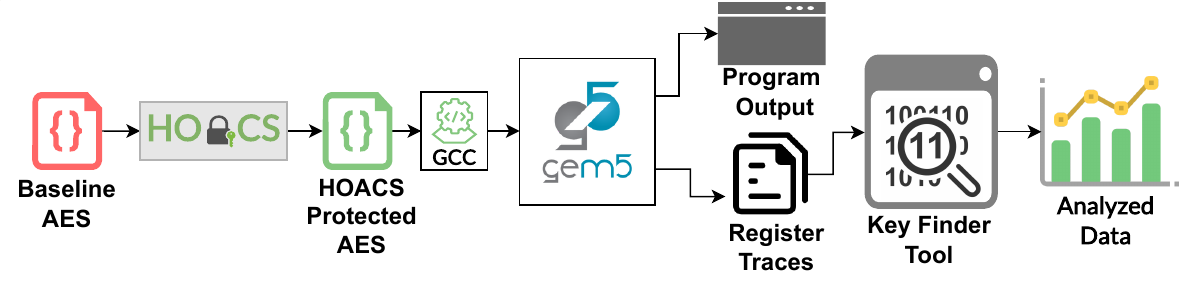}
    \caption{
        Experimental setup for integrating $HOACS-C$ into a baseline AES program and simulating it in the gem5 simulator with x86 architecture. Key Finder tool has been used to find secret assets in register traces and analyze them.
    }
    \label{exp_setup}
\end{figure} 
\section{Results}
\subsection{Security Analysis: Hardware Trojan Attacks}
\label{HT_Sec}
\subsubsection{Objective}
The primary objective of HOACS is to safeguard data from being leaked through CPU registers during a hardware Trojan attack. This experiment has been conducted to demonstrate the protection of sensitive data in registers using the HOACS framework.
\subsubsection{Experimental Setup}
\begin{figure*}[!t]
 \centering
 \includegraphics[width=0.95\linewidth]{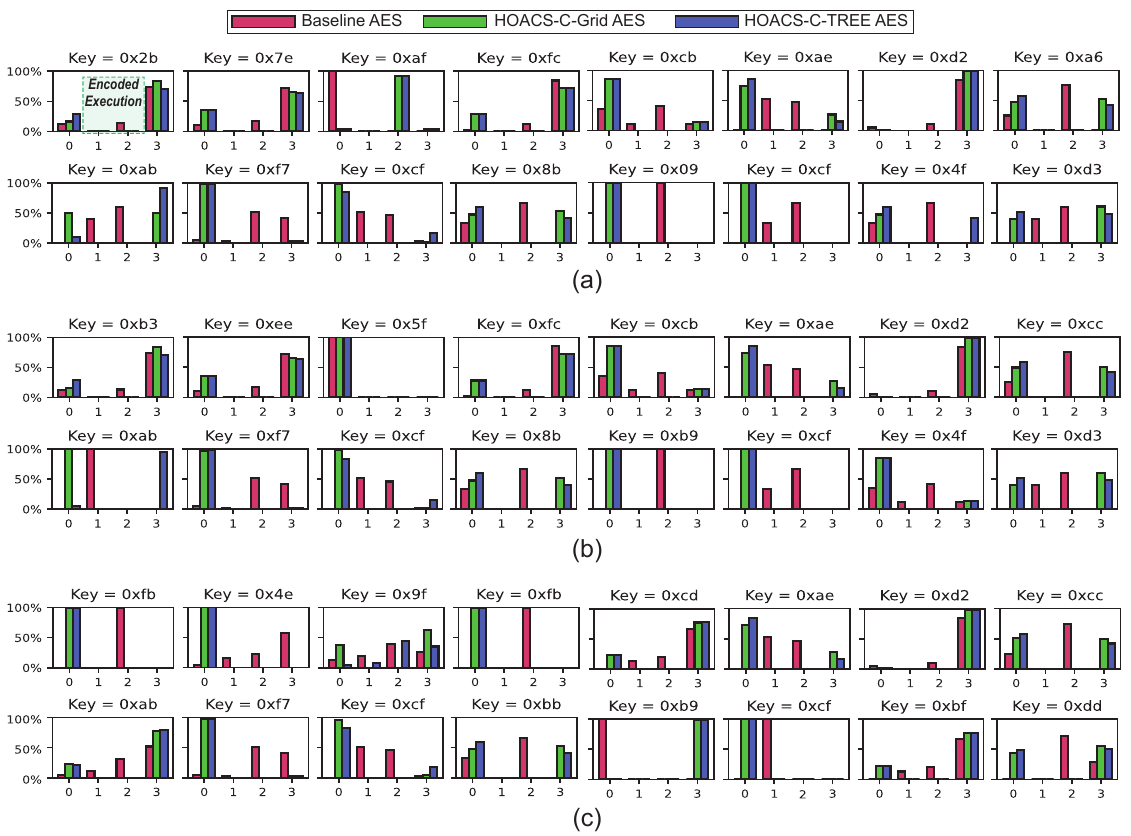}
 \caption{The percentage of key segments found in the registers during the Baseline AES, RNC-Tree, and RNC-Grid implementation executions. The X-axis denotes the four segmented parts of the whole program execution. RNC-encoded secure homomorphic execution is happening in segments 1 and 2, which is marked in the first subplot of this figure.}
 \label{x86_traces}
\end{figure*}

The $HOACS$-implemented AES and conventional AES programs have been tested on x86 architecture using the gem5 simulator to verify the security of the proposed method. The gem5 simulator offers time driven CPU with various debugging options \cite{gem5}. These debugging options include extracting physical addresses and register data, memory access, and the localization of executed instructions in every time cycle. Fig. \ref{exp_setup} illustrates the steps of our experimental setup and its flow. In our experiment, the baseline AES program was first converted into $HOACS$-protected AES using the proposed $HOACS-C$ tool. The GCC compiler was used to compile into binary without applying any optimization. Then the program binary was executed in gem5 using a time-driven simple CPU with 4GHz clock speed and 1GB of memory size. The gem5 debug flags enabled the dumping data of executed instructions and physical registers, which were divided into 4 parts for ease of analysis (0 to 3 in the subplots of Fig. \ref{x86_traces}). We used localization debug flags to localize the functions in terms of execution time. The four parts include the following functions of the AES program. \textbf{\emph{(a) Segment 0:}} initialization, storing plain key, plain texts, encoding the plain key. \textbf{\emph{(b) Segment 1:}} execution of key expansion function. \textbf{\emph{(c) Segment 2:}} generation of round keys. \textbf{\emph{(d) Segment 3:}} decoding the key and generating cipher text. 


\subsubsection{Analysis}
The proposed framework obfuscates the plain key into an RNC-encoded key. These RNC-encoded keys are then fed to the \emph{keyExpansion} module. Fig. \ref{x86_traces} shows our analysis for three different 128-bit (16 byte) keys below. 
\begin{align*}
key1 = 2b7eaffccbaed2a6abf7cf8b09cf4fd3 ~(Fig. \ref{x86_traces} (a)) \\
key2 = b3ee5ffccbaed2ccabf7cf8bb9cf4fd3 ~(Fig. \ref{x86_traces} (b)) \\
key3 = fb4e9ffbcbaed2ccabf7cfbbb9cfbfdd ~(Fig. \ref{x86_traces} (c))
\end{align*}
For each 16-byte key, we show our analysis separately for all 16 bytes. We searched the known key bytes in register data obtained from all four segments of the execution. As observed in all sub-figures, the plain key bytes can be found in all segments of baseline AES execution (red bars in different segments from 0 to 3). On the other hand, for proposed $HOACS$ tree and grid implementations, these plain keys are only found in the first($0$) and the last segment($3$) of the program execution when the key is encoded and decoded, respectively (green and blue bars). From key expansion (segment-$1$) to round key generation (segment-$2$), these original keys are absent in the CPU registers for $HOACS$-enabled AES for all 16 bytes of each of the three 128-bit keys being tested. This implies that during the key expansion and round key generation, the plain keys cannot be leaked using hardware Trojans.

We have observed some false positive results during this register data analysis. For example, in key-3, key byte 9f is found in segment-1 and segment-2 of the execution (Fig. \ref{x86_traces} (c)). The reasons for these false positive results are: (1) considering just a small 8-bit key segment and (2) frequent use of those numbers even in encoded values for other purposes. We further searched 9f during the data obtained in the first two keys and identified its presence, indicating that 9f appears in the registers irrespective of the key.     
\begin{figure*}[!t]
    \centering
    \includegraphics[width=0.95\linewidth]{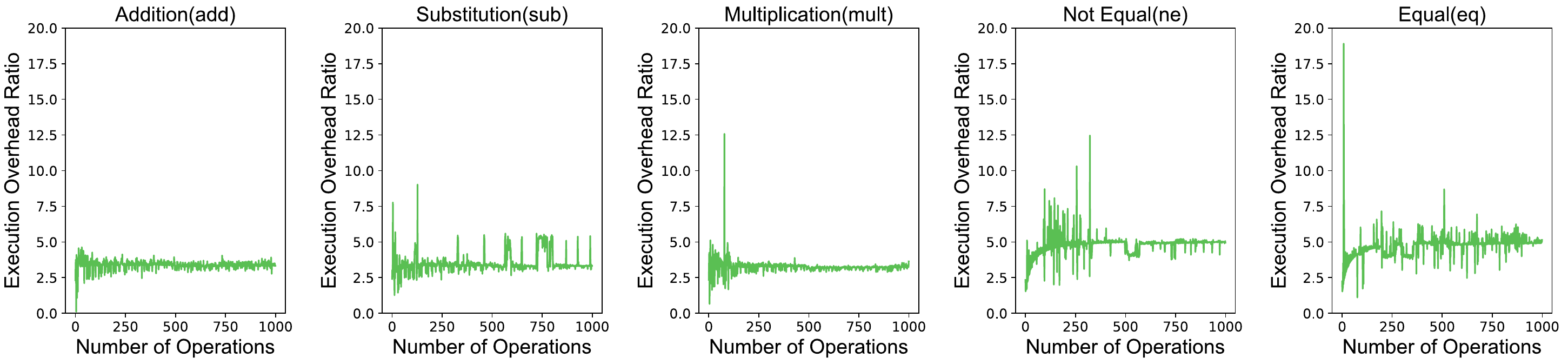}
    \caption{
        The execution overhead ratio for addition, multiplication, and logical operations.
    }
    \label{fig:ch3_compAdd}
\end{figure*} 

From Fig. \ref{x86_traces}, it has been observed that in the initial clock cycles, plain keys are visible in the registers. However, after initialization and encoding, the key bits become indistinguishable. Thus, a Trojan active during the initialization phase may leak the plain key. Nevertheless, Trojans are assumed to have hard-to-activate trigger conditions, making the probability of activation during the early stages of program execution very low. Consequently, even if the plain keys are exposed during the initial clock cycles prior to encoded execution, it is more likely that Trojans will be deactivated at that time. In Section \ref{subsec:trojan_feasibility}, we have provided an analysis of the Trojan activation time to understand the available cycles for encoding.
Exposing the data in the initialization stage can also be prevented if the confidential data is static and encoded during compilation. For this AES case study, the user can keep the plain key and $m$ values as static and encode them during compilation. Even though static data will be encoded during compilation, encoding of dynamic data (AES plain text) will not be affected because the $m$-values will be the same throughout the run-time. Our proposed \textit{HOACS-IR} implementation offers such encoding during compilation. 

\vspace{-0.35cm}
\subsection{Security Analysis: Unknown Moduli Attack}
\label{RNC_SEC}
\subsubsection{\textbf{Objective}}
In previous analyses, it has been demonstrated that HOACS can safeguard against the leakage of sensitive data through RNC encoding. However, what would occur if an attacker, utilizing a hardware Trojan attack, collects all the encoded data and attempts to decode it via a brute-force attack? This analysis presents the timing complexity an attacker would encounter should they attempt to decode the data by guessing the moduli.

\subsubsection{\textbf{Threat Model}}
Consider a scenario where an attacker tries to brute-force the actual number \( n \) without knowing the moduli \( m_1, m_2, \ldots, m_k \). The attacker is given the residues \( r_1, r_2, \ldots, r_k \).
\subsubsection{\textbf{Time Complexity of RNC to Binary Conversion}}
Given a system of \( k \) congruences:

\[
\begin{aligned}
    n &\equiv r_1 \, (\text{mod} \, m_1), \\
    n &\equiv r_2 \, (\text{mod} \, m_2), \\
    &\vdots \\
    n &\equiv r_k \, (\text{mod} \, m_k),
\end{aligned}
\]

where \( r_1, r_2, \ldots, r_k \) are residues and \( m_1, m_2, \ldots, m_k \) are pairwise co-prime moduli, the CRT guarantees a unique solution for \( n \) modulo \( M \), where \( M = m_1 \times m_2 \times \ldots \times m_k \).

The unique \( n \) can be computed as \cite{RNS_conversion}:

\[
n = \sum_{i=1}^{k} (r_i \times M_i \times y_i) \, (\text{mod} \, M),
\]

where \( M_i = M/m_i \) and \( y_i \) is the multiplicative inverse of \( M_i \) modulo \( m_i \).

Computing \( y_i \) for each \( i \) requires running the extended Euclidean algorithm, which has a time complexity of \( O(\log(m_i)^2) \). Given that there are \( k \) moduli, the total time complexity for finding all \( y_i \) and converting from the RNC representation back to the binary number representation is \( O(k \times \log(M)^2) \) \cite{EffectiveCRT}.

\subsubsection{\textbf{Complexity of Unknown Moduli Attack}}

The Unknown Moduli Attack's complexity can be delineated into two primary components: guessing the moduli and solving the congruences. The overall complexity, when combined, becomes computationally infeasible, as elaborated below.

\paragraph{Guessing the Moduli}
The first challenge is to identify the correct moduli. If we posit the maximum possible modulus as \( M \), due to the stipulation of coprimality between moduli, the total number of potential moduli sets exponentially grows in relation to \( k \). Mathematically, this translates to roughly \( M^k \).

\paragraph{Solving the Congruences}
Following the identification of a potential set of moduli, the next step involves solving the system of congruences. This computational step has a time complexity of \( O(k \times \log(M)^2) \).

\paragraph{\textbf{Overall Complexity}}
Merging both components' complexities, the total complexity of the attack is of the order \( O((M^k) \times k \times \log(M)^2) \). This magnitude of complexity signifies a significant computational challenge.

\paragraph{\textbf{Summary of the Complexity Analysis}} Given this level of complexity, it becomes evident that the task is practically infeasible. Utilizing even the most advanced computers available today, attempting a brute-force approach to discern the actual number \( n \) without prior knowledge of the moduli would necessitate an astronomically long duration.




\subsection{Performance Analysis}
To measure the overhead introduced by RNC operators, the execution time needed for different numbers of RNC operations was recorded. We used an Intel i7-9750H CPU clocked at around 4.1GHz with 16GB DDR4 RAMs for performance analysis. A number of(0 to 1000) instructions were executed to calculate the overhead of each arithmetic and logical operation. The process was repeated 30 times, and then the average was calculated. Each residue operation introduces some overhead, and the amount of resulting overhead depends on the complexity of the implementation of an operation. First, we calculated the execution time of regular compiled instructions to measure the overhead introduced by RNC operators. Then the process is repeated for RNC-Tree-based implementation, and the overhead ratios are shown in Fig. \ref{fig:ch3_compAdd}.

\begin{figure}[!h]
    \centering
    \includegraphics[width=0.95\linewidth]{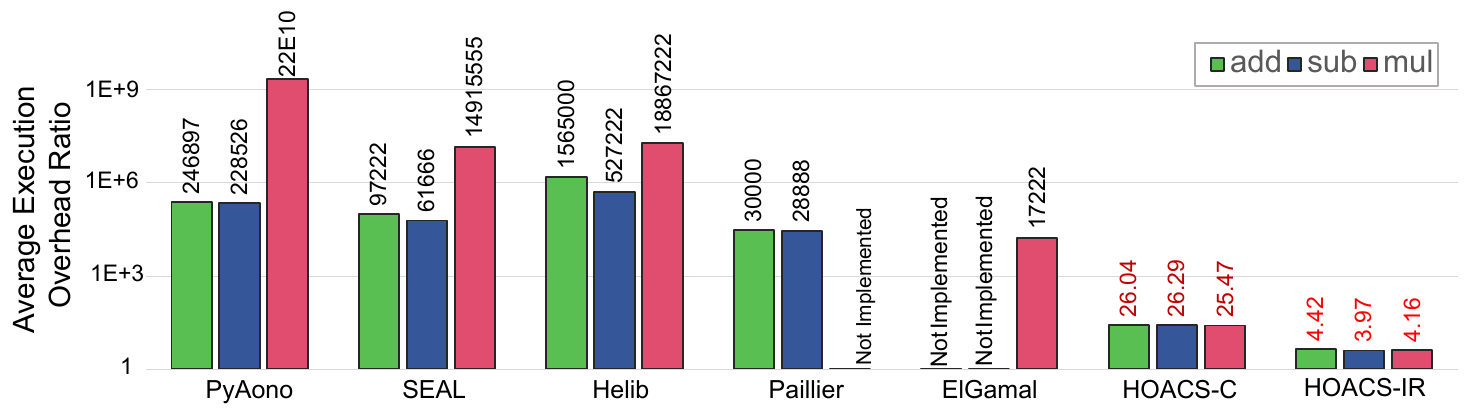} 
    \caption{
        Comparison of the average execution overhead ratio between existing homomorphic frameworks (PyAono\cite{pyaono}, SEAL\cite{sealcrypto}, Helib\cite{HELib}, Paillier \cite{comparison4}, ElGamal\cite{comparison4}) and our proposed $HOACS-C$ and $HOACS-IR$ frameworks. In this graph, the y-axis is plotted using a logarithmic scale.}
    \label{fig:comparison}

\end{figure} 

Inserting RNC protection at the LLVM IR level generally results in lower overhead than the default implementation of $HOACS-C$. Run-time randomization of data causes a substantial source of overhead, as shown in Table \ref{tab:overheadComparison}. When random multiple shifting is disabled ($HOACS-C$ W/out RAND of Table \ref{tab:overheadComparison}), the results show lower overhead. This randomization is needed to ensure more protection over the encoded data. For arithmetic operation in Table \ref{tab:overheadComparison},  randomization is enabled for $HOACS-C$ during encoding. Then, the random multiple shifts are removed prior to performing the desired operation. After the operation has been performed, random multiple shifts are added again if the output is an RNC-encoded value. These extra processes introduce overhead in arithmetic operations with $HOACS-C$ with random multiple shifting. No random shifting is done in the `equals' and `not equals' operations because these operations return a single boolean value. Thus these operations produce low overheads.
In $HOACS-IR$, applying RNC protections is part of the compiler process which allows the compile-time encoding of constants. Compile time encoding significantly reduces the run-time overhead.

In the comparative analysis of common arithmetic operations like addition, subtraction, and multiplication, this paper contrasts the \textit{HOACS-C} and \textit{HOACS-IR} methods with various homomorphic encryption schemes, as depicted in Figure \ref{fig:comparison}. Notably, our framework, grounded in RNC principles, exhibits significantly reduced overhead relative to frameworks based on traditional homomorphic encryption algorithms, such as PyAono, SEAL, Helib, Paillier, and ElGamal \cite{comparison4}. A key reason for this pronounced difference in overhead is attributed to the inherent characteristics of RNC. Unlike weighted number systems, RNC operates on integers by encoding them as sequences of remainders relative to a set of mutually coprime moduli. This unique approach facilitates parallel and autonomous processing of these remainders, significantly enhancing both speed and efficiency. Conversely, homomorphic encryption algorithms like PyAono, SEAL, Helib, Paillier, and ElGamal necessitate intricate arithmetic operations encompassing polynomial arithmetic, noise management, secure multi-party computation, and lattice-based computations. These operations are inherently more computationally intensive, contributing to the higher overhead observed in these systems.~\cite{gentry2009fully, cryptoeprint:2023/532}. The $HOACS$ framework, leveraging the RNC, specifically addresses data confidentiality concerns related to Trojan attacks without necessitating the encryption of the entire program. Instead, it focuses on encrypting only the sensitive portions of program execution. This targeted approach contrasts with other HE frameworks, which typically require the encryption of the whole program. While full-program encryption may offer enhanced confidentiality, it is generally superfluous for mitigating hardware Trojan threats. Hardware Trojans are primarily designed to intercept data, and the RNC-based framework is adept at safeguarding against such vulnerabilities. Given that the RNC foundation of the $HOACS$ framework substantially reduces computational overhead, it emerges as an ideal solution for systems with limited resources. 
The wider range of supported operations further extends the versatility of RNC-based protections.

\begin{table}[!t]
    \centering
     \caption{
    The average overhead ratio for each operation covered by both $HOACS-C$ and $HOACS-IR$ implementations}
    \resizebox{0.99\columnwidth}{!}{%
    \begin{tabular}{|c|m{1.5cm}|m{1.5cm}|c|}
        \hline
        Operator        & HOACS-C W/ Rand  & HOACS-C W/out Rand   & HOACS-IR    \\ \hline \hline
        Add             & 26.04             & 2.57                  & 4.42          \\ \hline 
        Sub             & 26.29             & 2.79                  & 3.97          \\ \hline 
        Multiplication  & 25.47             & 2.59                  & 4.16          \\ \hline 
        Equal           & 3.31              & 3.03                  & 5.78          \\ \hline 
        Not Equal       & 3.39              & 3.26                  & 5.72          \\ \hline
    \end{tabular}
   
    \label{tab:overheadComparison}
    }

\end{table}

\section{Discussion}
The overall results demonstrate that utilizing RNC for sensitive data protection offers a promising solution for COTS ICs. However, a pertinent question emerges: what will transpire if an attacker activates a Trojan during the encryption process? In the existing literature, other studies involving HE have regarded the encryption and decryption servers as secure, designating untrusted servers exclusively for HE-based computations \cite{HE_C1, HE_C2,IntelHEXLFPGA}. Contrarily, our proposed methodology conducts all operations — encryption, decryption, and computations — within the same untrusted COTS ICs. This approach affords a more robust solution for safeguarding sensitive data on COTS ICs.

\subsection{Feasibility of Trojan Attack During Encoding}
\label{subsec:trojan_feasibility}
If an attacker aims to leak the secret key during the encoding operation, they would need to activate a Trojan within that specific timeframe. This scenario necessitates the attacker to design a trigger circuit that is dependent on some function and data register. From the practical standpoint, we can assume that the Trojan will be activated if a certain data enters a specific register. However, designing a trigger based on this assumption poses a significant challenge for the attacker. This is because, even before the manufacturing stage, the attacker cannot predict which data register will be employed for the encoding operation. The feasibility analysis of two commonly used Trojan models have been shown in this paper. \textbf{Trojan Models}: \textbf{(a) Combinational Trojan}: This model is activated when specific conditions are met in a single clock cycle, such as when a particular value or a combination of values is present in the designated register or set of registers. \textbf{(b) Sequential Trojan}: This model is activated when a particular sequence of values is detected across multiple clock cycles. The register data can exhibit different values at each cycle, making the detection and activation of this Trojan more complex and stealthy.

It is pertinent to note that an attacker might design a Trojan that combines trigger inputs from multiple registers. However, discerning which combination of registers will contain the desired trigger data during the fabrication stage is arduous, even when the software program is known. This complexity arises as register allocation is often a dynamic process, influenced by various factors such as compiler optimization strategies and runtime conditions, thereby obfuscating predictable Trojan activation.

Considering these limitations of the attacker, let us assume that the attacker utilizes just one register and some instructions for activation. In this scenario, logic testing for Trojan activation can be executed during the IC testing stage. Logic testing involves verifying that the integrated circuit performs the desired functions in various conditions and does not malfunction under invalid inputs. The testing facility can perform an exhaustive test to activate such Trojan, which may potentially leak data through a data bus or memory alterations.

Exhaustive testing/verification involves applying all possible inputs to the IC, which is generally considered infeasible. However, our attack scenario limits the possible input space to test the specific Trojans under consideration (triggered with a combination of a register data and instruction). The instructions during the encoding phase are already known. Thus, we only need to test for all possible data combinations at each register, while the same encoding instructions execute.

The time required for this exhaustive testing/verification can be computed using the given formula for a combinational Trojan model where the register data is fixed:
\[
\text{Testing time (combinational)} = \sum_{0}^{N_{\text{DR}}} \left(\frac{2^b \times T_{\text{exec}}}{S_{\text{cpu}}}\right) 
\]

For a sequential Trojan model, assuming the trigger can assume a sequence of data value at each cycle, the formula might be:
\[
\text{Testing time (sequential)} = \sum_{0}^{N_{\text{DR}}} \left(\frac{2^b \times T_{\text{exec}}}{S_{\text{cpu}}} \right) \times \gamma 
\]

Where:
\begin{itemize}
    \item \( b \) is the size of the data register of the architecture.
    \item \( T_{\text{exec}} \) is the time taken for encoding operations.
    \item \( S_{\text{cpu}} \) is the speed of the CPU.
    \item \( N_{\text{DR}} \) is the number of data registers in the CPU.
    \item \( \gamma \) is the number of states in the sequential Trojan.
\end{itemize}

In our experiment using a \(4 \, \text{GHz}\) \texttt{x86-32bit CPU}, it takes \(66\) cycles to perform the encoding of the secret key. Using the given formula for the combinational Trojan, we find that the exhaustive testing/verification time for the combinational Trojan is about \(9.44\) minutes. Similarly, utilizing the formula for the sequential Trojan model and assuming \(\gamma = 5\), the exhaustive testing/verification time is approximately \(2.89\) minutes.

\subsection{Application of RNC in other algorithms }
The application of HOACS extends beyond merely safeguarding sensitive data. With the growing focus on machine learning and deep learning, concerns regarding data and model security have become increasingly prominent. To address these issues, privacy-preserving machine learning (PPML) has been introduced, utilizing fully homomorphic encryption (FHE) \cite{PPML1,PPML2,PPML3}. FHE ensures both data confidentiality and model privacy. Given that our proposed HOACS framework supports FHE with reduced overhead, our future research will aim to extend support for PPML within the HOACS framework.

\section{Conclusion}
This paper tries to address a significant security concern that has emerged due to the increased adoption of  COTS hardware. Even after more than a decade-long of research on hardware Trojans, trust issue in COTS components remains a critical challenge and has not received sufficient attention.   
Our proposed software-based solution \(HOACS\) provides confidentiality of secret assets at run time even in the presence of hardware Trojans in such untrusted COTS components. We have implemented \(HOACS\) in a practical case study of AES and performed extensive security analysis on the execution data for three different keys to confirm the confidentiality of the key during a target period. Additionally, we have proposed security analysis even if an attacker tries to brute-force the actual number without knowing the moduli and with given the residues. We also have integrated \(HOACS\) within a compiler tool-chain for automation and efficient integration. Our implementation shows promising performance overhead results compared to existing homomorphic solutions for data security. Future work will involve improving the run time performance further and identifying effective ways to mitigate the vulnerabilities during the initialization and encoding process. Systematic identification of the critical program regions for enabling the homomorphic operation can also be explored. Moreover, we discussed the feasibility of Trojan attacks during encoding for both combinational and sequential Trojan, which shows that it is easier for the logic testing facility to test and detect these kinds of Trojans, which are made specifically to leak secret data of certain programs.

\bibliographystyle{IEEEtran}
\bibliography{reference.bib}

\appendices
\setcounter{algorithm}{0} 
\renewcommand{\thealgorithm}{A\arabic{algorithm}} 

\section{Arithmetic Operation in RNC}

The algorithms of basic arithmetic operations have been provided in this appendix section. These algorithms have been used in the HOACS framework \cite{book1,book2,book3}.

\begin{algorithm}[H]
\caption{Addition in RNC}
\begin{algorithmic}[1]
\Statex \textbf{Input:} $A$, $B$ (operands), $\{m_1, m_2, \ldots, m_k\}$ (moduli)
\Statex \textbf{Output:} $C$ (result of $A + B$ in RNC)
\For{$i = 1$ to $k$}
    \State $A_i \gets A \mod m_i$
    \State $B_i \gets B \mod m_i$
    \State $C_i \gets (A_i + B_i) \mod m_i$
\EndFor
\State \Return $\{C_1, C_2, \ldots, C_k\}$
\end{algorithmic}
\end{algorithm}

\begin{algorithm}[H]
\caption{Subtraction in RNC}
\begin{algorithmic}[1]
\Statex \textbf{Input:} $A$, $B$ (operands), $\{m_1, m_2, \ldots, m_k\}$ (moduli)
\Statex \textbf{Output:} $C$ (result of $A - B$ in RNC)
\For{$i = 1$ to $k$}
    \State $A_i \gets A \mod m_i$
    \State $B_i \gets B \mod m_i$
    \State $C_i \gets (A_i - B_i + m_i) \mod m_i$
\EndFor
\State \Return $\{C_1, C_2, \ldots, C_k\}$
\end{algorithmic}
\end{algorithm}

\begin{algorithm}
\caption{Shifting in RNC}
\begin{algorithmic}[1]
\Statex \textbf{Input:} $A$ (operand), $n$ (shift amount), $\{m_1, m_2, \ldots, m_k\}$ (moduli)
\Statex \textbf{Output:} $D$ (result of shifting $A$ by $n$ in RNC)
\For{$i = 1$ to $k$}
    \State $A_i \gets A \mod m_i$
    \State $D_i \gets (A_i \times 2^n) \mod m_i$
\EndFor
\State \Return $\{D_1, D_2, \ldots, D_k\}$
\end{algorithmic}
\end{algorithm}

\begin{algorithm}[H]
\caption{Multiplication in RNC}
\begin{algorithmic}[1]
\Statex \textbf{Input:} $a$, $b$ (operands in RNC format), $\{m_1, m_2, \ldots, m_k\}$ (moduli)
\Statex \textbf{Output:} $result$ (result of $a \times b$ in RNC)
\State Initialize an empty list $result$
\For{$i = 1$ to $k$}
    \State $result_i \gets (a_i \times b_i) \mod m_i$
    \State Append $result_i$ to $result$
\EndFor
\State \Return $result$
\end{algorithmic}
\end{algorithm}

\begin{algorithm}[H]
\caption{Division in RNC}
\begin{algorithmic}[1]
\Statex \textbf{Input:} $a$, $b$ (operands in RNC format), $\{m_1, m_2, \ldots, m_k\}$ (moduli)
\Statex \textbf{Output:} $result$ (result of $a \div b$ in RNC)
\Function{ModInverse}{$x$, $m$}
    \For{$i = 1$ to $m - 1$}
        \If{$(x \times i) \mod m = 1$}
            \State \Return $i$
        \EndIf
    \EndFor
    \State \Return None \Comment{If no inverse exists}
\EndFunction
\State Initialize an empty list $result$
\For{$i = 1$ to $k$}
    \State $b\_inv \gets \Call{ModInverse}{b_i, m_i}$
    \If{$b\_inv$ is None}
        \State \textbf{raise} Error \Comment{"No modular inverse"}
    \EndIf
    \State $result_i \gets (a_i \times b\_inv) \mod m_i$
    \State Append $result_i$ to $result$
\EndFor
\State \Return $result$
\end{algorithmic}
\end{algorithm}

\begin{algorithm}[H]
\caption{Less-Than Comparison in RNC}
\begin{algorithmic}[1]
\Statex \textbf{Input:} $a$, $b$ (operands in RNC), $\{m_1, m_2, \ldots, m_k\}$ (moduli)
\Statex \textbf{Output:} Boolean result of $a < b$
\State $a_{std} \gets \text{ConvertToStandard}(a, \{m_1, m_2, \ldots, m_k\})$
\State $b_{std} \gets \text{ConvertToStandard}(b, \{m_1, m_2, \ldots, m_k\})$
\If{$a_{std} < b_{std}$}
    \State \Return True
\Else
    \State \Return False
\EndIf
\end{algorithmic}
\end{algorithm}

\begin{algorithm}[H]
\caption{Modulus Operation in RNC}
\begin{algorithmic}[1]
\Statex \textbf{Input:} $a$, $\{m_1, m_2, \ldots, m_k\}$ (moduli)
\Statex \textbf{Output:} $a$ in RNC
\For{$i = 1$ to $k$}
    \State $a_i \gets a \mod m_i$
\EndFor
\State \Return $\{a_1, a_2, \ldots, a_k\}$
\end{algorithmic}
\end{algorithm}

\begin{algorithm}[H]
\caption{Exponential Operation in RNC}
\begin{algorithmic}[1]
\Statex \textbf{Input:} $a$, $n$ (exponent), $\{m_1, m_2, \ldots, m_k\}$ (moduli)
\Statex \textbf{Output:} $a^n$ in RNC
\State Initialize $result$ as RNC representation of 1
\For{$i = 1$ to $n$}
    \State $result \gets \text{RNSMultiply}(result, a, \{m_1, m_2, \ldots, m_k\})$
\EndFor
\State \Return $result$
\end{algorithmic}
\end{algorithm}

\begin{algorithm}[H]
\caption{Bitwise OR Operation in RNC}
\begin{algorithmic}[1]
\Statex \textbf{Input:} $a$, $b$ (operands in RNC), $\{m_1, m_2, \ldots, m_k\}$ (moduli)
\Statex \textbf{Output:} Bitwise OR of $a$ and $b$ in RNC
\State $a_{bin} \gets \text{ConvertToBinary}(a, \{m_1, m_2, \ldots, m_k\})$
\State $b_{bin} \gets \text{ConvertToBinary}(b, \{m_1, m_2, \ldots, m_k\})$
\State $result_{bin} \gets a_{bin} \; \text{OR} \; b_{bin}$
\State $result \gets \text{ConvertToRNS}(result_{bin}, \{m_1, m_2, \ldots, m_k\})$
\State \Return $result$
\end{algorithmic}
\end{algorithm}

\begin{algorithm}[H]
\caption{Equality Operation in RNC}
\begin{algorithmic}[1]
\Statex \textbf{Input:} $a$, $b$ (operands in RNC format), $\{m_1, m_2, \ldots, m_k\}$ (moduli)
\Statex \textbf{Output:} Boolean result of $a = b$
\State // Step 1: Normalize operands by removing random multiple shifting
\State $a_{norm} \gets \text{RemoveShifting}(a, \{m_1, m_2, \ldots, m_k\})$
\State $b_{norm} \gets \text{RemoveShifting}(b, \{m_1, m_2, \ldots, m_k\})$
\State // Step 2: Compare RNC components
\For{$i = 1$ to $k$}
    \If{$a_{norm_i} \neq b_{norm_i}$}
        \State \Return False \Comment{Components are not equal}
    \EndIf
\EndFor
\State \Return True \Comment{All components are equal}
\end{algorithmic}
\end{algorithm}

\end{document}